\documentclass[conference]{IEEEtran}
\IEEEoverridecommandlockouts

\usepackage{graphicx}

\usepackage[linesnumbered,ruled,vlined,noend]{algorithm2e}
\usepackage{amsmath, amssymb, amsfonts}

\usepackage{amsthm}
\usepackage{subfig}
\usepackage{listings}
\usepackage{mathtools}

\usepackage{comment}

\newtheorem{theorem}{Theorem}
\newtheorem{lemma}{Lemma}
\newtheorem{definition}{Definition}
\newtheorem{corollary}{Corrollary}

\def\BibTeX{{\rm B\kern-.05em{\sc i\kern-.025em b}\kern-.08em
    T\kern-.1667em\lower.7ex\hbox{E}\kern-.125emX}}
\begin{document}

\title{Byzantine Fault-Tolerant Causal Ordering\\

}

\author{\IEEEauthorblockN{Anshuman Misra}
\IEEEauthorblockA{\textit{Department of Computer Science} \\
\textit{University of Illinois}\\
Chicago, USA \\
amisra7@uic.edu}
\and
\IEEEauthorblockN{Ajay Kshemkalyani}
\IEEEauthorblockA{\textit{Department of Computer Science} \\
\textit{University of Illinois}\\
Chicago, USA \\
ajay@uic.edu}
}

\maketitle

\begin{abstract}
Causal ordering in an asynchronous system has many applications in distributed computing, including in replicated databases and real-time collaborative software. Previous work in the area focused on ordering point-to-point messages in a fault-free setting, and on ordering broadcasts under various fault models. To the best of our knowledge, Byzantine fault-tolerant causal ordering has not been attempted for point-to-point communication in an asynchronous setting. In this paper, we first show that existing algorithms for causal ordering of point-to-point communication fail under Byzantine faults. We then prove that it is impossible to causally order messages under point-to-point communication in an asynchronous system with one or more Byzantine failures. We then present two algorithms that can causally order messages under Byzantine failures, where the network provides an upper bound on the message transmission time. The proofs of correctness for these algorithms show that it is possible to achieve causal ordering for point-to-point communication under a stronger asynchrony model where the network provides an upper bound on message transmission time. We also give extensions of our two algorithms for Byzantine fault-tolerant causal ordering of multicasts.
\end{abstract}


\begin{section}{Introduction}

Causality is an important tool in understanding and reasoning about distributed systems. 
However, determining causality in distributed systems is a challenging problem. This is due to the fact that there is no global physical clock to timestamp events as they occur. 
Theoretically causality is defined by the \emph{happens before} \cite{LLclock} relation on the set of events. 
In practice, logical clocks \cite{vc,vc2}, are used to timestamp events (messages as well) in order to capture causality.

If message $m1$ causally precedes $m2$ and both are sent to $p_i$, then $m1$ must be delivered before $m2$ at $p_i$ to enforce causal order. Causal ordering of messages is critical in applications that have multiple processes accessing common data. Causal ordering ensures that causally related updates to data occur in a valid manner respecting that causal relation. Applications of causal ordering include implementing distributed shared memory over message passing, fair resource allocation, collaborative applications such as real-time group editing of documents, event notification systems, and distributed virtual environments. A case can also be made for applying causal ordering to support or implement cryptocurrencies. 
 
 Most of the work on causal ordering of messages did not consider the Byzantine failure model. The only work on causal ordering under the Byzantine failure model was the recent result by Auvolat et al. \cite{auvolat2021byzantine} which considered only Byzantine-tolerant causal broadcasts, and the work in \cite{huang2021byz, kleppmann2020byzantine, 8935059} which relied on broadcasts. To the best of our knowledge, there has been no work on Byzantine-tolerant causal ordering of point-to-point messages. And as we show in this paper, the existing protocols for causal ordering of point-to-point messages and multicasts, such as \cite{birman1987reliable,DBLP:journals/dc/KshemkalyaniS98,DBLP:journals/jpdc/PrakashRS97,raynal1991causal,schiper1989new}, fail in the presence of Byzantine processes. It is important to solve this problem under the Byzantine failure model as opposed to a failure-free setting because it mirrors the real world. The Byzantine failure model in an asynchronous system deals with the most powerful adversary possible. If the problem is solvable under these circumstances then it will certainly be useful in practice.

 
The main contributions of this paper are as follows:

\begin{enumerate}
    \item The RST algorithm \cite{raynal1991causal} provides an abstraction of causal ordering of point-to-point and multicast messages, and all other (more efficient) algorithms can be cast in terms of this algorithm. We describe an attack called the artificial boosting attack that can force all communication to stop when running the RST algorithm. We show that one Byzantine node is enough to implement the attack. The artificial boosting attack is essentially a liveness attack. 
    
    
    \item We prove that causal ordering of point-to-point messages and multicasts in an asynchronous system with at least one Byzantine node is impossible.
    
    
    \item In view of the above impossibility result, we prove that a solution can be provided for causal ordering under a stronger asynchrony model. The strengthening is in the form of an upper bound on network transmission time.
    
    
    \item We propose two algorithms for causal ordering given an upper bound on network transmission time. These algorithms eliminate the $O(n^2)$ message space and time overhead of 
    \cite{birman1987reliable,DBLP:journals/dc/KshemkalyaniS98,DBLP:journals/jpdc/PrakashRS97,raynal1991causal,schiper1989new}, where $n$ is the number of processes in the system, and they use very small $O(1)$ control messages.
    
    
    \begin{enumerate}
        \item Sender-Inhibition Algorithm: This is simple to understand and implement. However send events at a process are blocking with respect to each other. This means that a process can initiate a message send only after the previous message has been received.
        
        
        \item Channel Sync Algorithm: This algorithm allows complete concurrency in the execution. 
        However, the implementation is a bit complicated -- it uses $n$ queues at each process.
    \end{enumerate}
     
    
    \item We give two algorithms implementing Byzantine fault-tolerant causal multicast in an asynchronous system by extending both the Sender-Inhibition algorithm and the Channel Sync algorithm. 
    
\end{enumerate}

\end{section}

\begin{section}{Previous Work}
\label{sec:previouswork}

Algorithms for causal ordering of point-to-point messages under a fault-free model have been described in \cite{schiper1989new,raynal1991causal}. 
These point-to-point causal ordering algorithms extend to implement causal multicasts in a failure-free setting \cite{DBLP:journals/dc/KshemkalyaniS98,DBLP:journals/jpdc/PrakashRS97}. The RST algorithm presented in \cite{raynal1991causal} is a canonical algorithm for causal ordering. We will analyze this algorithm and show that it does not work in a Byzantine failure setting. 

There has been significant work on causal broadcasts under various failure models. Causal ordering of broadcast messages under crash failures in asynchronous systems was introduced in \cite{birman1987reliable}. This algorithm required each message to carry the entire set of messages in its causal past as control information. The algorithm presented in \cite{mostefaoui2019crash} implements crash fault-tolerant causal broadcast in asynchronous systems with a focus on optimizing the amount of control information piggybacked on each message. An algorithm for causally ordering broadcast messages in an asynchronous system with Byzantine failures is proposed in \cite{auvolat2021byzantine}. The authors also made use of a result from \cite{guerraoui2019consensus} claiming that consensus does not need to be solved to implement a cryptocurrency and applied the Byzantine causal broadcast primitive to implement money transfer in a cryptocurrency. Despite causally ordering broadcasts in a Byzantine setting, this cannot be used to solve causal ordering of point-to-point messages in a Byzantine setting. This is because, this algorithm uses a vector clock based data structure to ensure that broadcasts are delivered in order. We show that a matrix clock based data structure is required for ordering point-to-point messages and therefore the broadcast technique will not be applicable to a point-to-point setting. There has been recent interest in applying the Byzantine fault model in implementing causal consistency in distributed shared memory and replicated databases \cite{8935059,huang2021byz,kleppmann2020byzantine}. In \cite{kleppmann2020byzantine}, Byzantine causal broadcast has been used to implement Byzantine eventual consistency. In \cite{8935059}, Byzantine reliable broadcast \cite{bracha1987asynchronous} is used to remove misinformation. induced by the combination of asynchrony and Byzantine behaviour. In \cite{huang2021byz}, PBFT (total order broadcast) \cite{DBLP:conf/osdi/CastroL99} is used to achieve consensus among non-Byzantine servers regarding the order of client requests.
To the best of our knowledge, no paper has attempted to solve causal ordering of point-to-point messages and multicasts in an asynchronous system with Byzantine failures. 

\end{section}

\begin{section}{System Model}
\label{sec:systemmodel}
The distributed system is modelled as a directed graph $G = (P,C)$. Here $P$ is the set of processes communicating asynchronously over a geographically dispersed network. $C$ is the set of communication channels over which processes communicate by message passing. The channels are assumed to be FIFO channels. $G$ is a complete graph with only one edge connecting each pair of processes. For a message send event $m$ at time $t_1$, the corresponding receive event is received at time $t_2 \in [t_1,\infty)$. A correct process behaves exactly as specified by the algorithm whereas a Byzantine process may exhibit arbitrary behaviour including crashing at any point during the execution. A Byzantine process cannot impersonate another process or spawn new processes.

Let $e^x_i$, where $x \geq 0$, denote the $x$-th event executed by process $p_i$. In order to deliver messages in causal order, we require a framework that captures causality as a partial order on a distributed execution. The \textit{happens before} \cite{LLclock} relation is an irreflexive, asymmetric, and transitive partial order defined over events in a distributed execution that captures causality. Happens before is denoted as "$\rightarrow$" and is defined as follows:

\begin{definition}

The happens before relation on events consists of the following rules:

\begin{enumerate}
    \item \textbf{Program Order}: For the sequence of events $\langle e_{i}^1, e_{i}^2, \ldots\rangle$ executed by process $p_i$, $\forall$ $k,j$ such that $k < j$ we have $e_{i}^k \rightarrow e_{i}^j$. 
    \item \textbf{Message Order}: If event $e_{i}^x$ is a message send event executed at process $p_i$ and $e_{j}^y$ is the corresponding message receive event at process $p_j$, then $e_{i}^x \rightarrow e_{j}^y$.
    \item \textbf{Transitive Order}: Given events $e$ and $e''$ in execution trace $\alpha$, if $\exists$ $e' \in \alpha$ such that $e \rightarrow e'$ $\land$ $e' \rightarrow e''$ then $e \rightarrow e''$.
\end{enumerate}

\end{definition}

Next, we define the happens before relation on messages.

\begin{definition} 
The happens before relation on messages is defined as follows:
\begin{enumerate}
\item The set of messages delivered from any $p_i \in P$ by a process is totally ordered by $\rightarrow$.
\item If $p_i$ sent or delivered message $m$ before sending message $m'$, then $m \rightarrow m'$.
\end{enumerate}
\end{definition}

We require an extension of the happens before relation to messages while accommodating the possibility of Byzantine behaviour. We present a partial order on messages called \textit{Byzantine happens before}.
Let $S$ be the set of all application-level messages delivered at correct processes in $P$.
The correctness of a point-to-point causal ordering algorithm in an asynchronous system with Byzantine failures is dependent on the definition of the partial order defined on $S$. This partial order is called the \textit{Byzantine happens before} relation and is denoted as $\xrightarrow{B}$. The Byzantine happens before relation is defined as follows:

\begin{definition}
The Byzantine happens before relation consists of the following rules:
\begin{enumerate}
    \item The set of messages delivered from any $p_i \in P$ by any correct process is totally ordered by $\xrightarrow{B}$. 
    
    
    \item If $p_i$ is a correct process and $p_i$ sent or delivered message $m$ (to/from another correct process) before sending message $m'$, then $m \xrightarrow{B} m'$.
\end{enumerate}
\end{definition}

The causal past of a message is defined as follows:

\begin{definition}

The \textit{causal past} of message $m$ is denoted as $P(m)$ and defined as the set of messages in $S$ that causally precede message $m$ in $\xrightarrow{B}$.

\end{definition}

We now define the correctness criteria that a causal ordering algorithm for point-to-point communication must satisfy.

\begin{definition}

A causal ordering algorithm for point-to-point messages must ensure the following:

\begin{enumerate}
    \item \textbf{Safety:} $\forall m' \in P(m)$ such that $m'$ and $m$ are sent to the same process, no correct process delivers $m$ before $m'$. 
    
    
    \item \textbf{Liveness:} All messages sent by non-faulty processes to non-faulty processes will be delivered eventually.

\end{enumerate}

\end{definition}

\end{section}

%
%
%

\begin{section}{Attacks due to Byzantine Behaviour}
\label{attacks}
All existing algorithms for implementing causal order for point-to-point messages in asynchronous systems use some form of \textit{logical timestamps}. This principle is abstracted by the RST algorithm \cite{raynal1991causal}.
Each message $m$ sent to $p_i$ is accompanied by a \textit{logical timestamp} in the form of a matrix clock providing information about send and receive events in the causal past of $m$, denoted as $P(send(m))$. This is to ensure that all messages $m^\prime$ $\in$ $P(send(m))$ whose destination is $p_i$ are delivered at $p_i$ before $m$. The implementation is as follows:

\begin{enumerate}
    \item Each process $p_i$ maintains (a) a vector $Delivered$ of size $n$ with $Delivered_i[j]$ storing a count of messages sent by $p_j$ and delivered by $p_i$, and (b) a matrix $M$ of size $n \times n$, where $M_{i}[j,k]$ stores the count of the number of messages sent by $p_j$ to $p_k$ as known to $p_i$. 
    
    
    \item When $p_i$ sends message $m$ to $p_j$, $m$ has a piggybacked matrix timestamp $M^m$, which is the value of $M_i$ before the send event. 
    Then $M_i[i,j] = M_i[i,j] + 1$.
    
    
    \item When message $m$ is received by $p_i$, it is not
    delivered until the following condition is met:
    
    \begin{center}
    
    $\forall k$, $M^m[k,i] \leq Delivered_i[k]$
        
    \end{center}
    
    
    \item After delivering a message $m$, $p_i$ merges the logical timestamp associated with $m$ with its own matrix clock, as $\forall j,k$, $M_i[j,k] = \max(M_i[j,k],M^m[j,k])$.
    
\end{enumerate}

A Byzantine process may fabricate values in the matrix timestamp in order to disrupt the causal ordering of messages in an asynchronous execution. The attacks are described in the following subsections.

\subsection{Artificial Boosting Attack}
\label{livenessattack}
A Byzantine process $p_j$ may increase values of $M_j[x,*]$ beyond the number of messages actually sent by process $x$ to one or more processes. When $p_j$ sends a message with such a Byzantine timestamp to any correct process $p_k$, it will result in $p_k$ recording Byzantine values in its $M_k$ matrix. These Byzantine values will get propagated across correct processes upon further message passing. This will finally result in correct processes no longer delivering messages from other correct processes because they will be waiting for messages to arrive that have never been sent.   

As an illustrative example, consider a system of $(n-1)$ correct  processes and a single malicious process -- $p_j$. At the time of executing an artificial boosting attack, $p_j$ forges values in its $M_j$ matrix as follows: if $p_j$ knows that $p_i$ (where $i$ may be $j$) has sent $x$ messages to $p_l$, it can set $M_j[i,l] = (x + d)$, $d >0$. When $p_k$ delivers a message from $p_j$, it sets $M_k[i,l] = (x + d)$. Finally, when $p_k$ sends a message $m$ to $p_l$, $p_l$ will wait for messages to arrive from $p_i$ (messages that $p_i$ has never sent) before delivering $m$. This is because $(Delivered_l[i] \leq x)$ $\land$ $(M^m[i,l] = x + d)$ $\Longrightarrow$ $(Delivered_l[i]  < M^m[i,l])$. Therefore, $p_l$ will never be able to deliver $m$. A single Byzantine process $p_j$ has effectively blocked all communication from $p_i$ to $p_l$. This attack can be replicated for all pairs of processes by $p_j$. 

Thus, a single Byzantine process can block all communication (including between each pair of correct processes), thus mounting a liveness attack. 
This liveness attack occurs under the $\rightarrow$ relation on messages and under the $\xrightarrow{B}$ relation on messages.

\subsection{Safety Violation Attack}
\label{safetyattack}
A Byzantine process $p_j$ may decrease values of $M^m[*,k]$ to smaller values than the true causal past of message $m$ and send it to a non-faulty process $p_k$. This may cause $m$ to get delivered out of order at $p_k$ resulting in a causal violation. Furthermore, if $p_j$ decreases the values of $M^m[*,*]$ to smaller values than the true causal past of message $m$ then, once $m$ is delivered to $p_k$ and $p_k$ sends a message $m'$ to correct process $p_l$, there may be a further causal violation due to a lack of transitive causal data transfer from $m$ to $p_k$ prior to event $send(m')$. These potential causal violations are a result of the possibility of a message getting delivered before messages in its causal past sent to a common destination.

As an illustrative example, consider a system of $(n-1)$ correct processes and a single malicious process -- $p_j$. At the time of executing a safety violation attack, $p_j$ forges values in its $M^m$ matrix as follows: if $p_j$ knows that $p_i$ has sent $x$ messages to $p_k$, $p_j$ can set $M^m[i,k] = x - 1$ and send $m$ to $p_k$. If $m$ is received at $p_k$ before the $x^{th}$ message $m'$ from $p_i$ is delivered, $m$ may get delivered before $m'$ resulting in a causal violation at $p_i$. 
In another attack, if $p_j$ knows that $p_i$ has sent $y$ messages to $p_l$, it can reduce $M^m[i,l] = y - 1$ and send $m$ to $p_k$. Assume $p_k$ delivers $m$ and sends $m'$ to $p_l$. If $m'$ arrives at $p_l$ before $m''$, the $y^{th}$ message from $p_i$ to $p_l$, arrives at $p_l$, $m'$ may get delivered before $m''$ resulting in a causal violation at $p_l$. In this way, a malicious process may cause causal violations at multiple correct processes by sending a single message with incorrect causal control information.

Note that safety is violated under the $\rightarrow$ relation on messages but not under the $\xrightarrow{B}$ relation on messages.

\end{section}

\begin{section}{Impossibility Proof}
\label{impossibility}
Causal order of messages can be enforced by either: (a) performing appropriate actions at the receiver's end, or (b) performing appropriate actions at the sender's end.

For enforcing causal ordering at the receiver's end, one needs to track causality in the distributed system, and some form of a logical clock is required to order messages (or events) by utilizing timestamps at the receiving process. Traditionally, logical clocks use transitively collected control information attached to each incoming message for this purpose. The RST abstraction \cite{raynal1991causal} described in Section~\ref{attacks} is used. 
However, in case there is a single Byzantine node $p_j$ in an asynchronous system, it can change the values  of $M_j$ at the time of sending $m$ to $p_i$. This may result in safety or liveness violations when $p_i$ communicates with a third process $p_k$ as explained in Section~\ref{attacks}. Lemma \ref{lemma:transitive} proves that transitively collected control information can lead to liveness attacks in asynchronous systems with Byzantine nodes.  Lemma~\ref{lemma:direct} then proves that even if the receiver $p_i$ tries to collect causal dependency information directly from all processes before delivering $m$, it is susceptible to liveness attacks. Lemma \ref{lemma:rec} combines the above two lemmas to show that no action by the receiver can guarantee the correctness of causal delivery of $m$.

As it is not possible to ensure causal delivery of messages by actions at the receiver's end, therefore, constraints on when the sending process can send messages need to be enforced to maintain causal delivery of messages. Each sender process would need to wait to get an acknowledgement from the receiver before sending the next message. Messages would get delivered in FIFO order at the receiver. While waiting for an acknowledgment, each process would continue to receive and deliver messages. This is important to maintain concurrency and avoid deadlocks. This can be implemented by using non-blocking synchronous sends, with the added constraint that all send events are \textit{atomic} with respect to each other.
However, Lemma \ref{lemma:snd} proves that even this approach would fail in the presence of one or more Byzantine nodes.  

Lemma \ref{lemma:bco} shows that the Byzantine happens before relation defined in Section~\ref{sec:systemmodel} cannot be tracked as a result of  Lemmas \ref{lemma:rec} and \ref{lemma:snd}. Theorem \ref{theorem:impossible} puts all these results together and proves that it is impossible to causally order messages in an asynchronous system with one or more Byzantine nodes.

\begin{lemma}
A single Byzantine process can execute a liveness attack when control information for causality tracking is transitively propagated and used by a receiving process for enforcing causal order.
\label{lemma:transitive}
\end{lemma}

\begin{proof}
Transitively propagated control information for causality tracking, whether by explicitly maintaining the counts of the number of messages sent between each process pair, or by maintaining causal barriers, or by encoding the dependency information optimally or by any other mechanism, can be abstracted by the causal ordering abstraction \cite{raynal1991causal}, described in Section~\ref{attacks}.
Each message $m$ sent to $p_k$ is accompanied with a \textit{logical timestamp} in the form of a matrix clock providing an encoding of $P(m)$. Based on the definition of $\xrightarrow{B}$, the encoding of $P(m)$ effectively maintains an entry to count the number of messages sent by $p_i$ to $p_j$, $\forall p_i,p_j \in P$. Such an encoding will consist of a total of $n^2$ entries, $n$ entries per process. Therefore, in order to ensure that all messages $m^\prime$ $\in$ $P(m)$ whose destination is $p_k$ are delivered at $p_k$ before $m$, the matrix clock $M$ whose definition and operation was reviewed in Section~\ref{attacks} is used to encode $P(m)$. 

Prior to delivering $m$ sent by $p_j$, $p_k$ will have to check that it is not merging Byzantine/malicious information from $M^m$ into $M_k$. 
In order to make sure that $p_j$ or its transitive predecessors along causal chains have not artificially increased $M^m[*,k]$ with the intention of executing an attack on liveness, $p_k$ will have to ask for and receive the current values of $M_i[i,k]$ from all $p_i$. This will put an upper bound on the amount of boosting that $p_j$ or its transitive predecessors could have done on $M^m[i,k]$ and prevent an attack on liveness because the boosting is limited to the actual messages that $p_i$ has sent. 
However, $p_i$ may never reply if it is Byzantine, and $p_k$ has no means of differentiating between a slow channel to/from a correct $p_i$ and a Byzantine $p_i$ that may never reply. So $p_k$ waits indefinitely.
Therefore the system is open to liveness attacks in the presence of a single Byzantine node. 
\end{proof}

\begin{lemma} \label{lemma:direct}
A single Byzantine process can execute a liveness attack when control information for causality tracking is directly obtained by a receiving process from the other processes for enforcing causal order.
\end{lemma}
\begin{proof}
When $p_k$ receives a message $m$ from $p_j$, $p_k$ can send a $probe$ message to each other process $p_i$ asking it to send back an {\em ack} on receipt of the {\em probe}. When $p_k$ receives the {\em ack} from $p_i$, $p_k$ can infer that all messages sent by $p_i$ to $p_k$ causally preceding $m$ have already been locally delivered because of flushing of the FIFO channels and local FIFO processing of message arrival queues. Therefore, after an {\em ack} is received from every such $p_i$, $m$ could be placed in the delivery queue for delivery. However, some $p_i$ that is Byzantine may never reply with sending the {\em ack}. $p_k$ has no way to differentiate between such a Byzantine $p_i$ and a correct $p_i$ to/from which the channel is very slow. So $p_k$ must keep waiting, and this is the liveness attack by $p_i$. 
\end{proof}

\begin{lemma}
\label{lemma:rec}
A single Byzantine process can execute a liveness attack when control information for causality tracking  is used by a receiving process for enforcing causal order.
\end{lemma}

\begin{proof}
Follows from Lemmas~\ref{lemma:transitive} and ~\ref{lemma:direct}.
\end{proof}

\begin{lemma}\label{lemma:snd}
A single Byzantine process can execute a liveness attack even if a sending process sends a message only when it is safe to send the message and hence its delivery at the receiver will not violate safety. 
\end{lemma}
\begin{proof}
The only way that a sending process $p_i$ can ensure safety of a message $m$ it sends to $p_j$ is to enforce that all messages $m'$ such that $m \xrightarrow{B} m'$ and $m'$ is sent to $p_j$ will reach the (common) destination $p_j$ after $m$ reaches $p_j$. Assuming FIFO delivery at a process based on the order of arrival, $m$ will be delivered before $m'$.

The only way the sender can enforce that $m'$ will arrive after $m$ at $p_j$ is not to send another message to any process $p_k$ after sending $m$ until $p_i$ knows that $m$ has arrived at $p_j$. $p_i$ can know $m$ has arrived at $p_j$ only when $p_j$ replies with an {\em ack} to $p_i$ and $p_i$ receives this {\em ack}. However, $p_i$ cannot differentiate between a malicious $p_j$ that never replies with the {\em ack} and a slow channel to/from  a correct process $p_j$. Thus, $p_i$ will wait indefinitely for the {\em ack} and not send any other message to any other process. This is a liveness attack by a Byzantine process $p_j$.
\end{proof}

\begin{lemma}
\label{lemma:bco}
The Byzantine happens before relation cannot be captured in an asynchronous system with one or more Byzantine processes.
\end{lemma}
\begin{proof}
From Lemma~\ref{lemma:rec}, neither directly obtained nor transitively obtained control information can be used for causally ordering messages at a receiver process in an asynchronous system with even one Byzantine process. 
From Lemma~\ref{lemma:snd}, it is not possible to causally order messages by introducing virtual synchrony by actions taken by a sender process in an asynchronous system with even one Byzantine process. Hence the Byzantine happens before relation cannot be captured in an asynchronous system  with even a single Byzantine process.
\end{proof}

\begin{theorem}
It is impossible to causally order point-to-point messages in an asynchronous message passing system with one or more Byzantine processes.
\label{theorem:impossible}
\end{theorem}

\begin{proof}
As a result of lemma \ref{lemma:bco}, we cannot capture the Byzantine happens before relation in an asynchronous system with one or more Byzantine nodes. It is not possible to introduce a causal ordering amongst a set of messages delivered at a process without a mechanism creating a relationship between messages ordered by some \textit{happens before relation}. Therefore, it becomes impossible to causally order messages in an asynchronous message passing system with one or more Byzantine processes. 
\end{proof}

\end{section}

\begin{section}{A Sender-Inhibition Algorithm}
\label{senderinhibition}
As a result of Theorem \ref{theorem:impossible}, we know that it is impossible to maintain both safety and liveness while trying to causally order messages in an asynchronous system with Byzantine faults. However, it is possible to extend the idea presented in Lemma \ref{lemma:snd} and develop a solution based on timeouts under a weaker asynchrony model. Under the assumption of a network guarantee of an upper bound $\delta$ on message transmission time, we prevent the Byzantine nodes from making non-faulty nodes wait indefinitely resulting in a liveness attack. This prevents a correct process from being unable to send messages because it is waiting for an acknowledgment from a Byzantine process. This solution can maintain both safety and liveness.

The solution is as follows: Each process maintains a FIFO queue, $Q$ and pushes messages as they arrive into $Q$. Whenever the application is ready to process a message, the algorithm pops a message from $Q$ and delivers it to the application. After pushing message $m$ into $Q$, each process sends an acknowledgement message to the sending process. Whenever process $p_i$ sends a message to process $p_j$, it waits for an acknowledgement to arrive from $p_j$ before sending another message. While waiting for $p_j$'s acknowledgement to arrive, $p_i$ can continue to receive and deliver messages. In case if $p_i$ does not receive $p_j$'s response within time $2*\delta$ (timeout period), it is certain that $p_j$ is faulty and $p_i$ can execute its next send event without violating $\xrightarrow{B}$.

\medskip

\begin{algorithm}

\SetAlgoLined
{\small
\SetKwComment{Comment}{$\triangleright$ }{}

\KwData{Each $p_i$ maintains a FIFO queue $Q$ and $lck$ is a lock common to all processes}

\medskip

\textbf{when} application is ready to process a message: \Comment{Deliver event} 

\Indp

$m = Q.pop()$ \\

\If{$m$ $\neq \phi$}{
deliver $m$

}

\Indm

\medskip

\textbf{when} message $m$ arrives from $p_j$: \Comment{Receive event} 

\Indp

$Q.push(m)$ \\

$send(ack,j)$ to $p_j$

\Indm

\medskip

\textbf{when} message $m$ is ready to be sent to $p_j$: \Comment{Send event} 

\Indp

$lck.acquire()$ \Comment{Executes atomically}

$send(m,j)$ to $p_j$ \\

start $timer$

\textbf{when} Acknowledgement arrives from $p_j$ for message $m$ $\lor$ timeout time exceeded

\Indp

$lck.release()$

} 
\caption{Sender-Inhibition Algorithm}
\label{alg:timeout}
\end{algorithm}

\medskip

Algorithm \ref{alg:timeout} consists of three \textit{when} blocks. The \textit{when} blocks execute asynchronously with respect to each other. This means that either the algorithm switches between the blocks in a fair manner or executes instances of the blocks concurrently via multithreading. In case a block has not completed executing and the process switches to another block, its context is saved and reloaded the next time it is scheduled for execution. If multithreading is used, each instance of a \textit{when} block spawns a unique thread. This maximizes the concurrency of the execution. Algorithm \ref{alg:timeout} ensures that while only one send event can execute at a given point in time, multiple deliver and multiple receive events can occur concurrently with a single send event.

\medskip

\begin{theorem}

\label{theorem:timeout}

Under a network guarantee of delivering messages within $\delta$ time, Algorithm \ref{alg:timeout} ensures liveness while maintaining safety.

\end{theorem}

\begin{proof}
The send event in Algorithm~\ref{alg:timeout} is implemented by the \textit{when} block in lines 8-13. A send event is initiated only after the previous send has released the lock, which happens when the sender $p_i$ (a) has received an {\em ack} from the receiver $p_j$, or (b) times out. 
\begin{enumerate}
\item In case (a), the sender learns that $p_j$ has queued its message $m$ in the delivery queue, and the sender can safely send other messages. Any message $m'$ such that $m \xrightarrow{B} m'$  and $m'$ is sent to $p_j$ will necessarily be queued after $m$ in $p_j$'s delivery queue. Due to FIFO withdrawal from the delivery queue, $m$ is delivered before $m'$ at $p_j$ and safety is guaranteed. As $p_i$ receives the {\em ack} before the timeout, progress occurs at $p_i$. There is no blocking condition for $m$ at $p_j$ and hence progress occurs at $p_j$.
\item In case (b) where a timeout occurs, the lock is released at $p_i$ and there is progress at $p_i$. It is left up to the application to decide how to proceed at $p_i$. This prevents a Byzantine process from executing a liveness attack by making a correct process wait indefinitely for the {\em ack}. It can be assumed that $p_j$ is a Byzantine process and so safety of delivery at $p_j$ does not matter under the $\xrightarrow{B}$ relation.
\end{enumerate}
Therefore, Algorithm~\ref{alg:timeout} ensures liveness while maintaining safety.
\end{proof}

In the Sender-Inhibition algorithm, the sender waits for at most $2*\delta$ time for the ack message to arrive from the receiver before sending its next message. The timeout period is fixed at $2*\delta$ because this is the maximum time an ack message can take to arrive from the point of sending the message.



\end{section}

\begin{section}{Channel Sync Algorithm}
\label{flushandsync}
In this section we present another solution to causally order messages in a system with Byzantine faults, utilizing timeouts. Similar to Algorithm~\ref{alg:timeout}, Algorithm~\ref{alg:timeout2} presents a solution that assumes that the underlying network guarantees that all messages are delivered within $\delta$ time. As long as this assumption holds, Algorithm~\ref{alg:timeout2} can guarantee both safety and liveness.  Each process maintains FIFO queues for each other process where it stores incoming messages from the concerned process. Application messages are delivered immediately after getting popped from the queue.  However, control messages are not processed immediately; the algorithm checks to make sure that it is safe to deliver the next message in the queue before completing processing. Whenever a process sends a message it informs every other process about the send event via a control message. Whenever a process delivers a message, it also informs every other process via a control message. Whenever process $p_i$ receives a control or application message from process $p_j$, it pushes it into $Q_j$. All control messages have timers associated with them to time them out in case of Byzantine behaviour of the sender and/or receiver. When $p_i$ pops a \textit{receive control message} from any queue $Q_x$ it waits for either the corresponding \textit{send control message} to reach the head of its queue (be dequeued), or the receive control message gets timed out in case the send control message does not arrive. This ensures that causality is not violated at $p_i$, while ensuring progress. We also need to ensure that in case of non-Byzantine behaviour on part of both the sender and receiver, both the send control message and receive control message do not time out before the other one arrives. In order to achieve this, the timer for receive control messages has to be set to at least $\delta$ as shown in Lemma~\ref{lemma:alg2_timer} while the timer for send control messages can be varied (see discussion below). The timer for send control messages can be reduced (it can be set to $0$ without compromising safety) to implement different behaviours in the system, but the timer for receive control message has to be at least $\delta$, and increasing it will only result in sub-optimal behaviour. Therefore, the timer for receive control messages should always be $\delta$.

\medskip

\medskip

\medskip
\begin{algorithm}
\SetAlgoLined
{\small
\SetKwComment{Comment}{$\triangleright$ }{}
\KwData{Each $p_i$ maintains a FIFO queue $Q_j$ for every process $p_j$}
\medskip
\textbf{when} the application is ready to send message $m$ to $p_j$: \\
\Indp
$send(m,j,app)$ to $p_j$ \\
\For{all $x \neq i,j$}{
$send(\langle i,j,sent \rangle, x, control)$ to $p_x$
}
\Indm
\medskip
\textbf{when} $\langle m,type \rangle$ arrives from $p_j$: \\
\Indp
$Q_j.push(m)$ \\
\If{$type = control$}{ 
start $timer$ for message $m$ \\
\If{$m[2] = sent$}{
\If{matching receive control message is in $Q_{m[1]}$ or popped}{
stop timers of send control message and matching receive control message
}
}
\If{$m[2] = delivered$}{
\If{matching send control message is in $Q_{m[1]}$ or popped}{
stop timers of receive control message and matching send control message
}
}
}
\Indm
\medskip
\textbf{when} the application is ready to process a message from $p_j$ and $\mid Q_j \mid \neq 0$: \hfill{\Comment{Only one instance of this block is executed at a time for a particular $Q_x$}}
\Indp
$\langle m,type \rangle = Q_j.pop()$ \\
\medskip
\If{$type = control$ $\land$ $m[2] = delivered$}{
 \While{timeout period not exceeded $\land$ timer not stopped}{wait in a non-blocking manner}
 \If{$timer$ is stopped}{\While{matching control message not reached head of $Q_{m[1]}$}{wait in non-blocking manner}
 }
 delete $m$
}
\medskip
\If{$type = control$ $\land$ $m[2] = sent$}{
 \While{timeout period not exceeded $\land$ timer not stopped 
 }
 {wait in a non-blocking manner
 }
 \If{timer stopped 
 }
 {delete the matching control message (popped/in $Q_{m[1]}$ if present)}
 delete $m$
}
\medskip
\If{$type = app$}{deliver $m$ \\
 \For{all $x \neq i,j$}{
 $send(\langle i,j,delivered \rangle, x,control)$ to $p_x$
 }
}
} 
\Indm
\caption{Channel Sync Algorithm}
\label{alg:timeout2}
\end{algorithm}

\begin{lemma}
Under the assumption of a network guarantee of delivering messages within a finite time period $\delta$, no receive control message with a timer greater than or equal to $\delta$ can get processed before the matching send control message.
\label{lemma:alg2_timer}
\end{lemma}

\begin{proof}
Without any loss of generality, we take $\delta_r = \delta$ and $\delta_s = 0$. Here $\delta_r$ and $\delta_s$ are timer wait times for receive control and send control messages, respectively. Whenever, a send control message arrives in Algorithm~\ref{alg:timeout2}, it stops the timer of the matching receive control message (if already present) to make sure that the receive control message waits for the send control message to get processed. If the send control message gets popped from the queue and the receive control message has not arrived, it simply gets processed. Now whenever the receive control message arrives, it waits for the timeout period and gets timed out without impacting safety because the send control message has already been processed.

In order to ensure that a receive control message waits for a send control message to get processed, we need to ensure that the send control message arrives before the receive control message times out. The maximum amount of time the send control message can take to arrive at any process $p_i$ is $\delta$ and the minimum amount of time the matching receive control message can take to arrive at $p_i$ is $0$. This means that in the worst-case scenario, the send control message will arrive in time $\delta$ after the arrival of the receive control message. Therefore, since the send control message arrives before the receive control message times out, the receive control message will have to wait for the matching send control message to get processed. (Note: the sender and receiver are non-Byzantine. If either of them is Byzantine, the receive control message, if present, will still time out at correct process $p_i$ but, as we will show in Corollary~\ref{corollary:alg2_live} and Theorem~\ref{theorem:alg2safety}, correctness of causal ordering is not impacted under $\xrightarrow{B}$.) 
\end{proof}

From Lemma~\ref{lemma:alg2_timer}, the timer for send control messages can be set as low as $0$ without impacting safety. The timer for send control messages can be tweaked based on the desired system performance. For instance, setting $\delta_s = 0$ would result in reduced latency for all send control messages at the expense of some receive control messages waiting out their entire waiting period of $\delta$ in the queue. If $\delta_s > 0$ a send control message waits after being popped until timeout. If in this interval any receive control message arrives, the receive control message gets deleted (lines 12-14, 27-28) and does not have to wait after being popped and until its timeout. So although the wait of a send control message increases, that of a receive control message decreases.
It would be interesting to simulate the effect on overall system latency by varying $\delta_s$ from $0$ upwards 
while keeping $\delta_r$ fixed at $\delta$ as per Lemma \ref{lemma:alg2_timer}.

If $\delta_s=0$ (effectively, no timer for send control messages), then in Algorithm~\ref{alg:timeout2}, stopping the send control message timer (lines 11,14) and testing if it was stopped (lines 25,27) can be replaced by setting and testing a boolean $flag\_timer\_stopped$.

A send event and a receive event are referred to as $s$ and $r$, respectively. The control messages we use for send and receive events are denoted $cms$ and $cmr$, respectively.

\begin{theorem}
\label{theorem:alg2bound}

Under the assumption of a network guarantee of delivering messages within a finite time period $\delta$, queued messages in Algorithm~\ref{alg:timeout2} will be dequeued in at most $max(\delta_s, \delta_r + max(\delta_s, \delta_r))$ time.

\end{theorem}

\begin{proof}

As a simplifying assumption, the time taken to pop a message from a queue is considered to be $0$. The time each message spends in the queue is only because of latency induced by control messages. Let $m$ be an application message inserted in  $Q_{i_0}$ at process $p_j$ at time 0 (as a reference instant). The waiting time in the queue can be analyzed as follows.

\begin{enumerate}
\item There may be no control messages in front of $m$ in $Q_{i_0}$. Since the latency induced by application messages that may be in front of $m$ is $0$, $m$ will be popped and delivered immediately. The waiting time in the queue for $m$ is $0$.
    
    
\item There may be one or more send control messages before $m$ in $Q_{i_0}$. Each of the control messages will take at most $\delta_s$ time to get processed. Since the timers for all of those control messages are ticking concurrently, $m$ will have to wait for at most $\delta_s$ time.
    
    
\item There may be one receive control message $cmr_{i_0}$ in front of $m$ in $Q_{i_0}$. $cmr_{i_0}$ is for application message $m_1$ sent from $i_i$ (before time 0) to $i_0$ (received before time 0). Note, if there are multiple receive control messages ahead, the analysis can be independently made for each of them. 
\begin{enumerate}
    \item $cms_{i_1}$ does not arrive in $\delta_r$. $cmr_{i_0}$ times out at $\delta_r$. So total delay is $\delta_r$.
    \item Otherwise $cms_{i_1}$ is inserted in $Q_{i_1}$ in time $\delta_r$ from time 0. 
    \begin{enumerate}
        \item It may be blocked by $cms'_{i_1}$. This times out in $\delta_s$ time. Total delay is therefore $\delta_r + \delta_s$.
        \item It may be blocked by $cmr_{i_1}$ for application message $m_2$ from $i_2$ sent before time 0 to $i_1$ received before time 0, ahead in $Q_{i_1}$. Therefore $cmr_{i_1}$ arrived within time $\delta_r$ from time 0. It waits for $cms_{i_2}$.
    \end{enumerate} 
\end{enumerate}

\item Reasoning for the delay introduced by wait for $cms_{i_2}$, corresponding to application message $m_2$, in $Q_{i_2}$ is as follows.
\begin{enumerate}
    \item $cms_{i_2}$ does not arrive in $\delta_r$. $cmr_{i_1}$ times out in $\delta_r$ after its arrival which was latest at $\delta_r$ from time 0. Total delay is therefore $\delta_r + \delta_r$.
    \item Otherwise $cms_{i_2}$ arrived within $\delta_r$ from time 0 because $m_2$ was sent before time 0 due to transitive chain $m_2 \rightarrow m_1$ and $m_1$ was received before time 0. Therefore $cms_{i_2}$ is inserted in $Q_{i_2}$ in $\delta_r$ from time 0.
    \begin{enumerate}
        \item It may be blocked by $cms'_{i_2}$. This times out in $\delta_s$ time. Total delay is therefore $\delta_r + \delta_s$.
        \item It may be blocked by $cmr_{i_2}$ for application message $m_3$ from $i_3$ sent before time 0 to $i_2$ received before time 0, ahead in $Q_{i_2}$. Therefore $cmr_{i_2}$ arrived within time $\delta_r$ from time 0. It waits for $cms_{i_3}$.
    \end{enumerate}
\end{enumerate}

\item The reasoning for the delay introduced by wait for $cms_{i_3}$ in $Q_{i_3}$is identical to the reasoning for the wait introduced by $cms_{i_2}$ in the previous item. In particular, $cms_{i_3}$ was inserted in $Q_{i_3}$ within $\delta_r$ from time 0.
\end{enumerate}

We generalize the above analysis as follows. Define $\leftarrow$ as the ``waits for" or ``succeeds in time" relation on control messages in the queues at $P_j$. Then, there exists a chain of control messages
\[
cmr_{i_0} \leftarrow cms_{i_1} \leftarrow cmr_{i_1} \leftarrow cms_{i_2} \leftarrow cmr_{i_2} \leftarrow 
\ldots \leftarrow cms_{i_k}
\]
each of which must have arrived in the corresponding $Q_{i_\alpha}$ within time $\delta_r$ from time 0 (see ($\ast$) below). This chain corresponds to the following chain of application messages:
\[m_k \rightarrow m_{k-1} \rightarrow \ldots m_2 \rightarrow m_1
\]

We prove that 
``($\ast$) $cmr_{i_{a-1}}$ is inserted in $Q_{i_{a-1}}$  within time $\delta_r$ from time 0, $cms_{i_a}$ was inserted in $Q_{i_a}$ within time $\delta_r$ from time 0." 
We use induction. The base case, being for $a=2$, was shown above. Assume the induction hypothesis is true for $x, x\geq 2$. We show the result ($\ast$) for $x+1$.
As $cmr_{i_{x}}$ arrives in $Q_{i_{x}}$ before $cms_{i_{x}}$, from the induction hypothesis for $x$, $cmr_{i_x}$ is inserted in $Q_{i_x}$ within $\delta_r$ from time 0. It waits for $cms_{i_{x+1}}$.
    $cms_{i_{x+1}}$ arrived within $\delta_r$ from time 0, because $m_{x+1}$ was sent before time 0 due to transitive chain $m_{x+1} \rightarrow m_x \rightarrow \ldots m_1$ and $m_1$ was received before time 0 (because $cmr_{i_0}$ was received in $Q_{i_0}$ before time 0). Therefore $cms_{i_{x+1}}$ is inserted in $Q_{i_{x+1}}$ within $\delta_r$ from time 0.
(end of proof of ($\ast$))

We also claim $k$ is finite and bounded because the corresponding control messages existed in the queues at $P_{i_0}$ only at time 0 or later and were therefore added to the queues at the earliest at $- \max(\delta_r,\delta_s)$; this implies the corresponding application messages were therefore sent after $- \delta - \max(\delta_r,\delta_s)$.   

The chain of control messages terminates at $cms_{i_k}$, for $k>0$, as analyzed by the following cases.
\begin{enumerate}
    \item There is nothing ahead of it in $Q_{i_k}$. Total delay this queue contributes is $\delta_r$. Total overall delay contributed by queues $Q_{i_1}$ to $Q_{i_k}$ combined is as per (2)-(4) below.
    \item There are only send control messages ahead of it in $Q_{i_k}$, and they time out. Total delay contributed by this queue is $\delta_r + \delta_s$. This is also the overall total combined delay of $m$ contributed by queues $Q_{i_1}$ through $Q_{i_k}$. 
    \item There are only receive control messages ahead of it in $Q_{i_k}$ and they time out. Total delay this queue contributes is $\delta_r + \delta_r$. If in all queues $Q_{i_{\alpha}}$, $1 \leq \alpha \leq k$, there are no send control messages $cms'_{i_{\alpha}}$ ahead of $cms_{i_{\alpha}}$, total overall combined delay for $m$ these queues combined contribute is also $\delta_r + \delta_r$, otherwise it is $\delta_r + \max(\delta_r + \delta_s)$.
    \item There are send and receive control messages ahead of it in $Q_{i_k}$ and they time out. Total delay is $\delta_r + \max(\delta_r, \delta_s)$. This is also the total overall combined delay contributed by queues $Q_{i_1}$ through $Q_{i_k}$ combined.
    \item If there is a send control message ahead of $m$ in $Q_{i_0}$, this queue contributes a delay of $\delta_s$. Total overall delay contributed by all queues $Q_{i_0}$ to $Q_{i_k}$ is $\max(\delta_s,z)$, where $z$ is the total combined delay contributed by queues $Q_{i_1}$ to $Q_{i_k}$ as analyzed in the above cases. 
\end{enumerate}
If $k=0$, there is no receive control message ahead of $m$ in $Q_{i_0}$, and as shown at the start of the proof, total delay is bounded by $\delta_s$. Combining all these, the total overall combined delay of $m$ is bounded by $\max(\delta_s, \delta_r + \max(\delta_r + \delta_s))$.

    
    

\end{proof}

Since the amount of time each message spends in the message queue is bounded by a finite quantity, every application message will eventually be delivered. Therefore liveness is maintained by Algorithm~\ref{alg:timeout2}.

\begin{corollary}

\label{corollary:alg2_live}

Algorithm \ref{alg:timeout2} guarantees liveness. 

\end{corollary}

\begin{theorem}

\label{theorem:alg2safety}

Under the assumption of a network guarantee of delivering messages within a finite time period $\delta$, Algorithm~\ref{alg:timeout2} can guarantee safety by setting timers for control messages as a function of 
$\delta$.
\end{theorem}

\begin{proof}

In order to ensure safety, prior to delivering any message $m'$ at process $p_j$, we need to ensure that if $\exists m \in P(m')$ such that $m$ is sent to $p_j$, then $m$ is delivered before $m'$ at $p_j$. 

Algorithm~\ref{alg:timeout2} ensures safety at any process as follows:

\begin{itemize}
    \item \textbf{Program Order:} Since we assume FIFO channels, messages from $p_i$ to $p_j$ get enqueued in $Q_i$ in program order and get delivered in program order.
    
   \medskip
    
    \item \textbf{Transitive Order:} Let $m$ be sent by $p_i$ to $p_j$ at send event $s_i^x$. Consider a causal chain of $b$ messages starting at $s_i^y$ from $i = i_0$ and $j = i_b$ as : 
    
    \begin{center}
        $\langle s_i^y = s_{i_0} \rightarrow r_{i_1} \rightarrow s_{i_1} \rightarrow r_{i_2} \rightarrow .... \rightarrow r_{i_{b-1}} \rightarrow s_{i_{b-1}} \rightarrow r_{i_b} \rangle$
    \end{center}
    
    Let $s_i^x \xrightarrow{B} s_i^y$ and $m'$ be the last message of causal chain $\langle s_{i_{b-1}} \rightarrow r_{i_b} \rangle$. This means that $m \in P(m')$ by transitivity. 
    
    Based on the following observations at $p_j$:
    
    \begin{enumerate}
        \item In $Q_{i_0}$, $cms_{i_0}$ (control message for $s_{i_0}$) waits for $m$ (sent at $s^x_{i_0}$) to get delivered.
        
        \item From Lemma~\ref{lemma:alg2_timer}, in $Q_{i_{\alpha}}$ ($1 \leq \alpha \leq (b-1)$), $cmr_{i_{\alpha}}$ waits for $cms_{i_{\alpha - 1}}$ in $Q_{i_{\alpha -1}}$ to be processed. 
        
        \item In $Q_{i_{\alpha}}$ ($1 \leq \alpha \leq (b-2)$), $cms_{i_{\alpha}}$ waits for $cmr_{i_{\alpha}}$ to be processed. 
        
        \item In $Q_{i_{b-1}}$, $m'$ (sent at $s_{i_{b-1}}$) waits for $cmr_{i_{b-1}}$ to be processed.
    \end{enumerate}
    
    Hence, message $m'$ waits for message $m$ to get delivered.
      
\end{itemize}

Algorithm \ref{alg:timeout2} therefore ensures safety:  ``that $\forall m \in P(m')$ sent to the same $p_j$, $m$ gets delivered before $m'$ at $p_j$," under a network guarantee of delivering messages within a fixed time. 
\end{proof}

\end{section}

\begin{section}{Adaptations to Multicast}
\label{multicast}
In a multicast, a send event sends a message to multiple destinations that form a subset of the process set $P$. Different send events by the same process can be addressed to different subsets of $P$. This models dynamically changing multicast groups and dynamic membership in multicast groups. There can exist overlapping multicast groups. In the general case, there are $2^{|P|} -1$ groups. Although there are several algorithms for causal ordering of messages under dynamic groups, such as \cite{DBLP:journals/dc/KshemkalyaniS98,DBLP:journals/jpdc/PrakashRS97}, 
none of them consider the Byzantine failure model.
All the existing algorithms use transitively collected control information about causal dependencies in the past -- they vary in the size of the control information, whether in the form of causal barriers as in \cite{DBLP:journals/jpdc/PrakashRS97} or in the optimal encoding of the theoretically minimal control information as in 
\cite{DBLP:journals/dc/KshemkalyaniS98}. The RST algorithm still serves as a canonical algorithm for the causal ordering of multicasts, and it can be seen that the same liveness attack described in Section~\ref{attacks} can be mounted on these algorithms. Furthermore, the same impossibility result of Theorem~\ref{theorem:impossible} along with its correctness proof carries over to the multicast setting.

The Sender-Inhibition algorithm and the Channel Sync algorithm also work for causal ordering of multicast messages in the face of Byzantine failures, under the assumption of the network guarantee of an upper bound $\delta$ on the message transmission time. The modifications to adapt these algorithms are given next.

\subsection{Adaptation of the Sender-Inhibition Algorithm}
Make the following changes to the Sender-Inhibition algorithm.
\begin{enumerate}
\item Line 8 would change to: ``{\bf when} $m$ is ready to be sent to group $G$:"

\item Line 10 would change to: ``$send(m,j)$ to each $p_j \in G$"

\item Line 12 would change to: ``{\bf when} Acknowledgement arrives from each $p_j \in G$ for message $m$ $\lor$ timeout time exceeded"
\end{enumerate}

\subsection{Adaptation of the Channel Sync Algorithm}
The major changes to Algorithm~\ref{alg:timeout2} to get Algorithm~\ref{alg:timeout2m} are as follows. (1) The send control message contains the group members instead of the receiver $j$ in the second parameter (line 4). (2) When this control message is received, the parameter $G$ is manipulated to track the matching receive control messages in their queues (lines 10,13,16). (3) When popped, a send control message deletes the matching receive control messages (lines 29-30).
The proofs of correctness (safety and liveness) are almost identical to those for Algorithm~\ref{alg:timeout2}.

As in Algorithm~\ref{alg:timeout2}, $\delta_r=\delta$, while $\delta_s$ can be 0 or larger with the same trade-offs. If $\delta_s=0$ (effectively no timer for send control messages), receive control messages in their queues may wait $\delta_r$ until they time out as they may not get deleted when the send control message gets popped and deleted (because the receive control message arrived after that time). 

It is not necessary to send the group members $G$ in the second parameter of the send control message and we can eliminate this space overhead and corresponding time overhead for processing $G$. In this case, stopping the send control message timer is not useful (because we cannot track the matched receive control messages in order to delete them (lines 29-30)) nor is it possible (lines 13,16). This implies that a send control message must not need a timer (i.e., $\delta_s$ is effectively set to 0), and the send control message's second parameter $G$ is to be replaced by $x$ in line 4. Correctness of the algorithm is not impacted.

\medskip
\begin{algorithm}[t]
\SetAlgoLined
{\small
\SetKwComment{Comment}{$\triangleright$ }{}
\KwData{Each $p_i$ maintains a FIFO queue $Q_j$ for every process $p_j$}
\medskip
\textbf{when} the application is ready to send message $m$ to group $G$: \\
\Indp
$send(m,j,app)$ to each $p_j \in G$\\
\For{all $x \neq i$}{
$send(\langle i,G,sent \rangle, x, control)$ to $p_x$
}
\Indm
\medskip
\textbf{when} $\langle m,type \rangle$ arrives from $p_j$: \\
\Indp
$Q_j.push(m)$ \\
\If{$type = control$}{ 
start $timer$ for message $m$ \\
\If{$m[2] = sent$}{
 $D = m[1]$ \\
 \For{all $x \in D$}{
  \If{matching receive control message is in $Q_{x}$ or popped}{
   stop timer of matching receive control message; $D = D \setminus \{x\}$; stop timer if $D = \emptyset$
  }
 }
}
\If{$m[2] = delivered$}{
 \If{matching send control message $c$ is in $Q_{m[1]}$ or popped}{
 stop timer; $c.D = c.D \setminus \{m[0]\}$; stop timer of $c$ if $c.D = \emptyset$
 }
}
}
\Indm
\medskip
\textbf{when} the application is ready to process a message from $p_j$ and $\mid Q_j \mid \neq 0$: \hfill{\Comment{Only one instance of this block is executed at a time for a particular $Q_x$}}
\Indp
$\langle m,type \rangle = Q_j.pop()$ \\
\medskip
\If{$type = control$ $\land$ $m[2] = delivered$}{
 \While{timeout period not exceeded $\land$ timer not stopped}{wait in a non-blocking manner}
 \If{$timer$ is stopped}{\While{matching control message not reached head of $Q_{m[1]}$}{wait in non-blocking manner}
 }
 delete $m$
}
\medskip
\If{$type = control$ $\land$ $m[2] = sent$}{
 \While{timeout period not exceeded $\land$ timer not stopped 
 }
 {wait in a non-blocking manner
 }
 \For{all $x \in m[1] \setminus D$}{
  delete the matching control message (popped/in $Q_x$ if present)
 }
 delete $m$
}
\medskip
\If{$type = app$}{deliver $m$ \\
 \For{all $x \neq j$}{
 $send(\langle i,j,delivered \rangle, x,control)$ to $p_x$
 }
}
} 
\Indm
\caption{Channel Sync Algorithm for Multicast}
\label{alg:timeout2m}
\end{algorithm}

\end{section}

\begin{section}{Discussion}

It has been proven that Byzantine causal broadcast is solvable  \cite{auvolat2021byzantine}. Therefore, an important question arises -- Why is Byzantine fault-tolerant causal broadcast achievable whereas Byzantine fault-tolerant causal order for point-to-point communication impossible 
in asynchronous systems? 
From the impossibility result of Theorem~\ref{theorem:impossible}, the problem is that a single Byzantine adversary can launch a liveness attack by artificial boosting. In Byzantine causal broadcast, all messages are sent to every process in the system and the underlying Byzantine reliable broadcast layer \cite{bracha1987asynchronous} ensures that every correct process receives the exact same set of messages. 
Upon receiving $m$, the receiving process simply waits for its logical clock to catch up with $m$'s timestamp (each broadcast delivered will increment one entry in the logical clock) and deliver $m$ once it is safe to do so. After delivering message $m$, the receiving processes' logical clock is greater than or equal to $m$'s timestamp. This means that the receiving process does not need to merge message $m$'s timestamp into its own logical clock upon delivering $m$. Since there is no logical clock merge operation after receiving a message, no amount of artificial boosting can result in a liveness attack in Byzantine causal broadcast. In case of causal ordering in point-to-point communication, every process receives a different set of messages. When a process $p_i$ delivers a message, it means that $p_i$ has delivered all messages addressed to it in the causal past of $m$. However, it requires the timestamp attached to $m$ to ascertain the messages in the causal past of $m$ that are not addressed to $p_i$. Therefore, the receiving process needs to merge the timestamp of the delivered message into its own logical clock so that subsequent messages sent by it can be timestamped with their causal past. 

We presented two algorithms for causal ordering under a stronger asynchrony model. The Sender-Inhibition algorithm has reduced concurrency because each sender has to wait for its message to be received before sending the next message. However it is very easy to implement. The Channel Sync algorithm 
does not inhibit concurrency (beyond what is necessary to enforce causal order) but its implementation is complicated.
%
%

A potential application of the Channel Sync algorithm is in implementing cryptocurrencies. As shown by \cite{guerraoui2019consensus}, weaker alternatives to consensus can be used to prevent \textit{double spending} in the money transfer problem. Preventing double spending is the core problem being solved by cryptocurrencies \cite{nakamoto2008bitcoin}. The Channel Sync algorithm can be used to order all transactions that have dependencies with each other. Intuitively, all transactions from a single user will be ordered by the Byzantine happens before relation preventing double spending attacks. The control messages will have to be augmented to store additional information which will be used by the receiving process to add transactions to its ledger. The Byzantine causal broadcast in \cite{auvolat2021byzantine} inherits the constraint of $f \leq n/3$, (where $f$ is the number of Byzantine processes and $n$ is the total number of processes) from the underlying Byzantine reliable broadcast layer. However, the Channel Sync algorithm does not have any such constraint. In the worst case where $f = (n-2)$, the remaining two correct processes will be able to communicate with safety and liveness guarantees. Therefore, investigating the potential of causal order via the Channel Sync algorithm in solving the money transfer problem is an interesting and important area of future work.

\end{section}

\begin{section}{Conclusion}

This paper gave a formal definition of Byzantine causal order and demonstrated a liveness attack on the canonical technique to implement causal order in an asynchronous system with Byzantine faults.  We proved that it is impossible to implement Byzantine fault-tolerant causal order for point-to-point messages in an asynchronous system due to the possibility of liveness attacks. We then showed that it is possible to implement Byzantine fault-tolerant causal order under a stronger asynchrony model. The Sender-Inhibition algorithm and the Channel Sync algorithm were presented to implement causal order under a network guarantee of an upper bound on message transmission time. 
These algorithms were extended to implement causal multicast. 

One future area of work is to investigate other possible strengthenings of the asynchrony model under which causal order is solvable and provide formal solutions for the same. An immediate area of interest is to find applications for both the Sender-Inhibition algorithm and the Channel Sync algorithm. It would be especially interesting to see whether a cryptocurrency can be implemented with causal order using the Channel Sync algorithm instead of using consensus.

\end{section}

\bibliographystyle{IEEEtran}
\bibliography{references}

\end{document}